# Experimental Investigation of the Structural Performance of Composite Structures Produced using Additive Manufacturing

Hunter Watts[1], Ryan Premo[2], Wayne Huberty[3], Chris Bounds[4], and Han-Gyu Kim[5]

*Mississippi State University, Mississippi State, MS 39762, USA*

**This project is focused on investigating the structural performance of parts and structures produced using the latest additive manufacturing techniques. For additive manufacturing of test coupons, fused deposition modeling of fiber-reinforced polymer (RFP) and high-resolution low-force stereolithography (LFS) thermoset resin printing systems were employed. UV thermoset resin was used for LFS printing, while RFP printing adopted two different types of filaments: amorphous polycarbonate carbon fiber filaments and semi-crystalline Nylon 12 glass fiber filaments. For the experimental work, specimens were printed for tension, compression, and shear tests. Additionally, mode-II interlminar fracture in these specimens was explored. The elastic modulus and strength values of these specimens were compared with the data of oven-cured T700G/2510 composites. The experimental work herein will be extended to develop damage models for 3D-printed structural parts and structures for aerospace and space applications.**

## I. Introduction

Additive manufacturing has evolved to become more affordable, varied, and sophisticated and will play a more important role in the aerospace industry and space programs. The first Federal Aviation Administration (FAA) certification of 3D-printed structural titanium components in Boeing 787 [1] and NASA's On-Orbit Servicing, Assembly, and Manufacturing 2 project [2] are representative examples. The damage mechanisms and structural integrity of these parts and structures, however, have not been rigorously covered in the literature. To address this issue, this project will focus on investigating the structural performance of composites produced using the latest additive manufacturing methods for aerospace and space applications. For additive manufacturing of composite specimens, Fused Deposition Modeling (FDM) of Fiber-Reinforced Polymer (FRP) and high-resolution Low Force Stereolithography (LFS) thermoset resin printing systems were employed. First, the thermoset high-resolution LFS system can produce smooth and continuous surfaces for efficient aerodynamics without postprocessing while utilizing the cross-linking of polymers for higher strength and thermo-tolerance compared to thermoplastics [3,4], which is required for high-speed vehicles. Second, the FRP printing system uses thermoplastic resin with discontinuous fibers that provide increased layer adhesion and stiffness. The experimental work herein is focused on obtaining the in-plane mechanical properties of 3D-printed specimens from tensile, compression, and shear tests and characterizing mode-II interlaminar fracture in these specimens through End-Notched Flexure (ENF) tests. Comparing the mechanical properties of the 3D-printed specimens with conventional prepreg cases and understanding the quasi-brittle damage process in these structures are of primary interest. The mode-II fracture case was chosen as it is more relevant to high-speed aircraft applications than the mode-I and mixed mode-I/II cases [5].

---

[1] Research Engineer I, Advanced Composites Institute.
[2] Graduate Research Assistant, Department of Aerospace Engineering.
[3] Associate Director, Advanced Composites Institute.
[4] Director, Advanced Composites Institute.
[5] Assistant Professor, Department of Aerospace Engineering. Corresponding author (hkim@ae.msstate.edu).

## II. Specimen Design and Manufacturing

### A. Specimen details

The test matrix of this work is presented in Table 1. For tension, compression, and shear tests, in-plane specimens were designed and manufactured based on the ASTM specifications [6-8]; however, the thickness dimensions of the specimens were larger than the specifications due to the inherent characteristics of additive manufacturing. Similarly, the ENF specimens followed the ASTM specifications for mode-II interlaminar fracture [9] with larger thickness values. The manual recommends using non-adhesive film inserts (≤ 13 µm thick) to induce a pre-crack along the midplane at one end; however, due to the difficulties in embedding an insert during the additive manufacturing process, end-notches were instead printed during the additive manufacturing process. The notch thickness was 0.2 mm, which was significantly thicker than the ASTM specification. Thus, stress concentrations around the notch tip were expected to occur during the fracture process in the ENF specimens.

**Table 1. Test matrix**

| Test type | Property type | ASTM specification |
| --- | --- | --- |
| Tension tests in the print direction | Tensile modulus and strength | D3039/D3039M-17 [6] |
| Compression tests in the print direction | Compressive modulus and strength | D6641/D6641M-16e2 [7] |
| V-notched beam tests under shear | 1-2 shear modulus and strength | D5379/D5379M-19e1 [8] |
| Mode-II interlaminar fracture | Mode-II interlaminar fracture toughness | D7905/D7905M-19e1 [9] |

### B. Additive manufacturing details

For the additive manufacturing of the specimens, two different types of printers were employed as shown in Fig. 1. For high-resolution LFS printing, a Formlabs Form 3+ system (see Fig. 1a) was employed. This system has a 300-µm layer resolution utilizing a 250-mW laser. LFS works by using a vat of resin, a build plate, and a precision laser to create a part. For each layer, the build plate is lowered into the vat of resin until there is only a layer's worth of resin between the build plate and the bottom of the clear vat. Subsequently, a laser that resides underneath the vat moves along a track from left to right rapidly firing to cure the resin inside the vat in predetermined areas. The bottom of the vat is made from a flexible material so that the laser can apply a small force against the build plate as it moves for better layer adhesion. Once the layer is complete, the build plate will rise out of the resin vat to allow the excess resin to drain off the build plate and allow a stirring device to travel through the vat of resin, ensuring that each layer is consistently mixed. Once the print is complete, the specimens are removed from the build plate and placed in a vat of 99% isopropyl alcohol and mechanically stirred to remove any excess resin from the specimens. Once the specimens are cleaned, they are left to air dry before they are placed into a UV curing oven. The cure oven emits 405 nm light while also heating the chamber up to 80° C depending on the resin used. On the other hand, a Bambu Lab X1-Carbon (see Fig. 1b) was employed for FRP printing. The system is equipped with a 7 µm-resolution Lidar camera and a 500

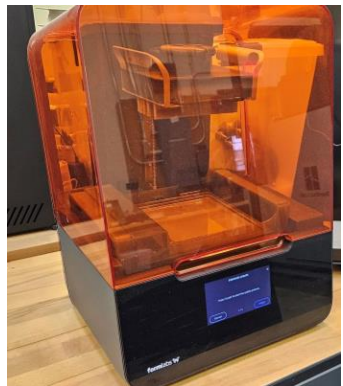   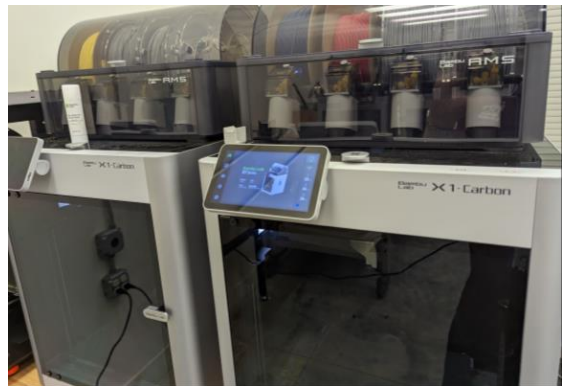

(a)                                                                 (b)

**Figure 1. Additive manufacturing systems. (a) Formlabs Form 3+ system. (b) Bambu Lab X1-carbon system.**

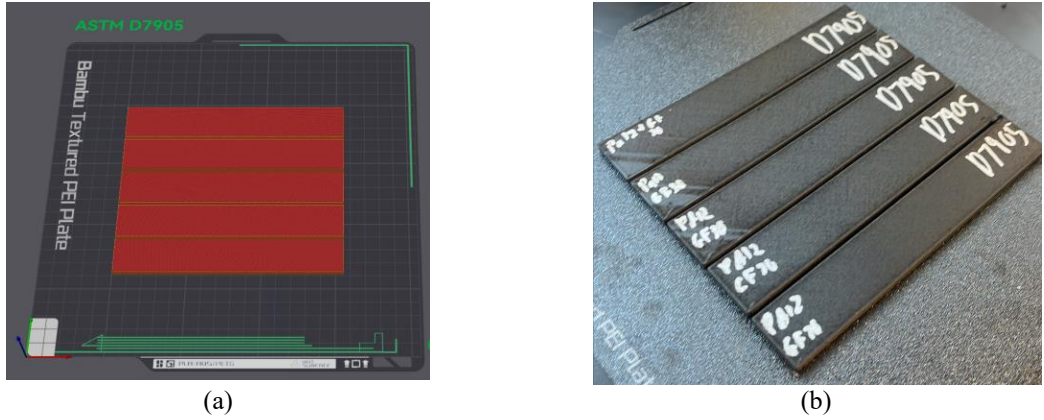

(a)                                                     (b)

**Figure 2. Additive manufacturing of ENF specimens using the Bambu Lab X1-carbon system. (a) A software preview of ENF specimens showing the print orientation and plane. (b) The manufactured specimens immediately after printing.**

mm/s- speed extruder for repeatable high-quality and high-speed printing. All the FDM printed specimens were printed in the ±45˚ print direction on the XY plane as shown in Fig. 2. Lastly, three different types of engineering-grade materials were selected for additive manufacturing considering their wide applications and availability in the industry as shown in Table 2. Each of the fiber-loaded thermoplastic materials was properly dried according to the manufacturer's recommendations in the technical data sheet [11,12], and all materials used were stored in a humidity-monitored and climate-controlled environment between printings.

**Table 2. Additive manufacturing materials**

| Product | Material type | Manufacturing system | Manufacturing type |
|---|---|---|---|
| Formlabs Tough 1500 resin [10] | UV thermoset resin | Formlabs Form 3+ system | LFS printing |
| 3DXTECH CarbonX™ ezPC+CF printing filament [11] | Amorphous polycarbonate carbon fiber (CF-PC) filament (100% infill) | Bambu Lab X1-carbon system | RFP printing |
| 3DXTECH FibreX™ PA12+GF30 glass fiber reinforced PA12 3D filament [12] | Semi-crystalline Nylon 12 glass fiber (PA12-GF30) filament (100% infill) | Bambu Lab X1-carbon system | RFP printing |

## III. Experimental Setup

For multi-scale experimentation, two types of digital image correlation (DIC) systems were employed as shown in Fig. 3. A macroscopic DIC system (see Fig. 3a) provided coupon-scale data by capturing the stress-strain development on the entire specimen area. A Shimadzu AGX-V load frame and a Correlated Solutions VIC 2D package were employed for the coupon-scale tests. A microscopic DIC system (see Fig. 3b), on the other hand, was used to characterize fracture processes occurring at small strains in specimens such as the mode-II interlaminar tests. For the microscopic tests, a Psylotech μTS testing frame, an Olympus BXFM microscope system with a 12-MP machine vision camera, and the VIC 2D package were employed. The FOV of the microscopic testing was 10 mm x 7.3 mm, and the lens was focused on the notch tips (i.e., the initial crack tips) to capture crack propagation along the fracture process zones.

For the V-notched beam tests, the microscopic system was initially applied to obtain high-resolution data. The FOV available with the microscope, however, was not large enough to capture both V notches on the beams and shear failure could unpredictably be initiated from one of the notches. To address this issue, a hybrid experimental setup was built by merging the aspect of the macroscopic testing system with the μTS testing frame as shown in Fig. 4. In this testing configuration (see Fig. 4a), the machine vision camera was paired with the macroscopic camera lens instead of the microscope and was cantilevered over the shear test fixtures. This combination allowed the complete capture of the coupon surface. A DIC image of a speckled V-notched coupon in the test fixture shown in Fig. 4b was captured immediately preceding the shear fracture. The purple contours indicate large strains propagated between the notches.

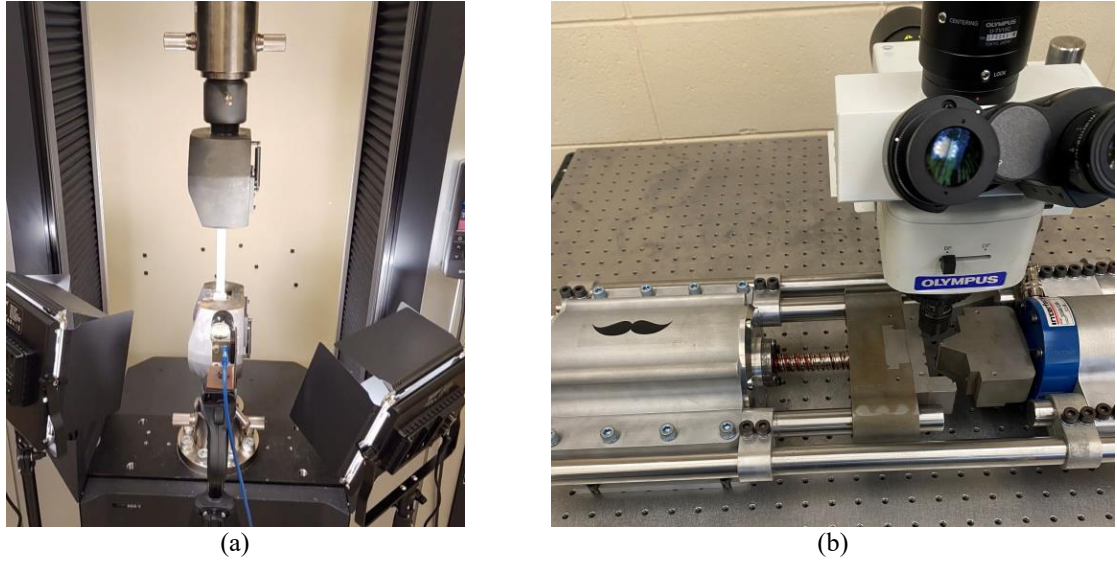

(a)                          (b)

**Figure 3. Experimental setup for multi-scale DIC tests. (a) A Shimadzu AGX-V load frame for macroscopic DIC tests. (b) A Psylotech µTS testing frame for microscopic DIC tests.**

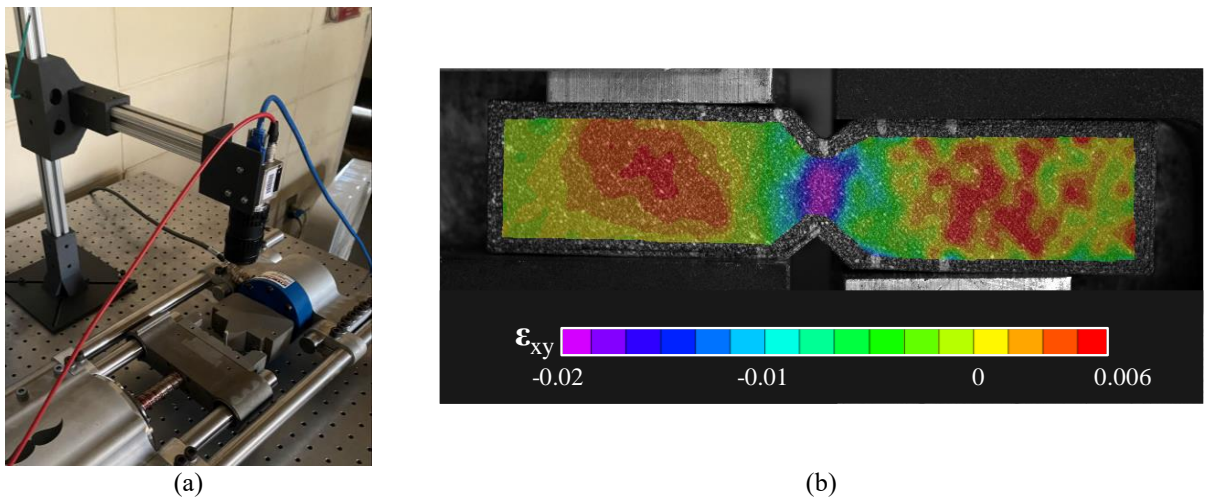

(a)                          (b)

**Figure 4. V-notched beam test of a CF-PC specimen in a Psylotech µTS testing frame. (a) A hybrid test setup. (b) DIC analysis result of the image for shear strain.**

## IV. Experimental Results

### A. In-plane material properties

The experimental data sets for the in-plane tests were analyzed to obtain the elastic modulus and strength values of the materials under tension, compression, and shear loads. Five tests were conducted within each test set (see Table 1) for each material (see Table 22), and the results were averaged to acquire single material properties. The stress-strain curves are presented in Fig. 5. In this paper, the subscripts *xx*, *yy*, and *xy* denote the stress or strain for the longitudinal, transverse, and shear properties, respectively. The stress-strain curves obtained from the experimental data are presented in Figs. 5 to 7. The specimens all showed consistent response under tension (see Fig. 5); however, the resin specimens (see Fig. 5a) showed lots of oscillations on the curves. Additionally, large plastic behaviors were observed from the resin and CF-PC specimens (see Figs. 5a and 5b, respectively) under tension. The compression curves (see Fig. 6) showed slightly more inconsistency in the curves compared to the tension cases. Furthermore, less plastic response was observed from the compression tests compared to the tension sets. Lastly, the V-notched beam tests (see Fig. 7) showed very inconsistent response for the CF-PC and PA12-F30 specimens (see Figs. 7b and 7c, respectively). This was caused by the shear fracture process in these specimens at low strain levels.

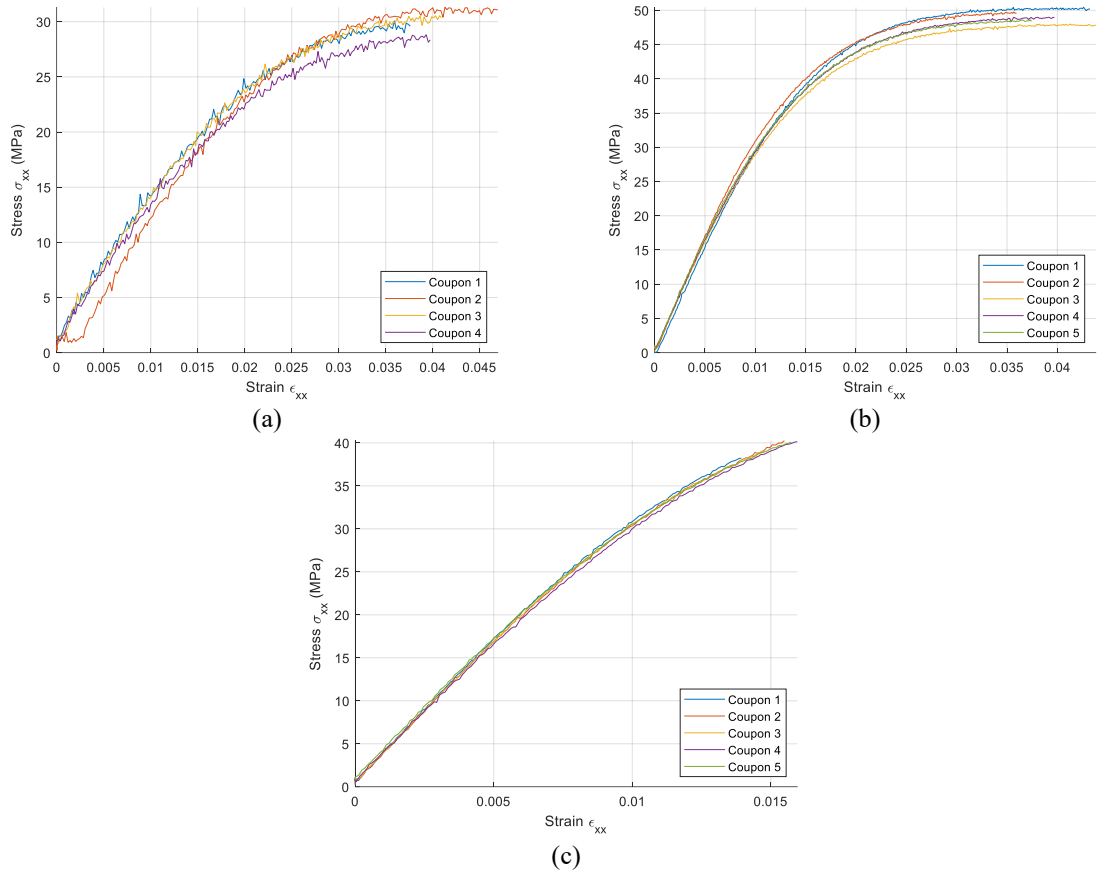

**Figure 5. Full-scale stress-strain curves for the tension tests. (a) Resin. (b) CF-PC. (c) PA12-F30.**

The elastic modulus and strength values of these specimens are tabulated in Tables 3 and 4. In these tables, comparisons are made using the data of oven-cured Toray T700G/2510 composites [13]. The 3D printed specimens showed significantly lower modulus and strength values compared to the 0° properties of the oven-cured composites. The shear and 90° values of the T700G/2510 composites were relatively comparable to the corresponding values of

**Table 3. Test summary and comparison with the 0° properties**

| Test types | | Modulus (GPa) | | | Strength (MPa) | | |
|---|---|---|---|---|---|---|---|
| | | Test | T700G/2510* | % difference | Test | T700G/2510* | % difference |
| **Tension tests** | Resin | 1.0523 | 125 | -99.2 | 30.7818 | 2172 | -98.6 |
| | CF-PC | 2.9224 | | -97.7 | 49.2500 | | -97.7 |
| | PA12-F30 | 2.9744 | | -97.6 | 39.5960 | | -98.2 |
| **Compression tests** | Resin | 1.5123 | 112 | -98.7 | 39.0269 | 1448 | -97.3 |
| | CF-PC | 2.7679 | | -97.5 | 53.4882 | | -96.3 |
| | PA12-F30 | 3.0830 | | -97.3 | 62.4211 | | -95.7 |
| **V-notched beam tests** | Resin | 1.6853 | 4.23 | -60.2 | 20.9098 | 86.2 | -75.7 |
| | CF-PC | 3.1128 | | -26.4 | 37.5711 | | -56.4 |
| | PA12-F30 | 2.5141 | | -40.6 | 30.4770 | | -64.6 |

*The values were taken from the manufacturer's report [13].

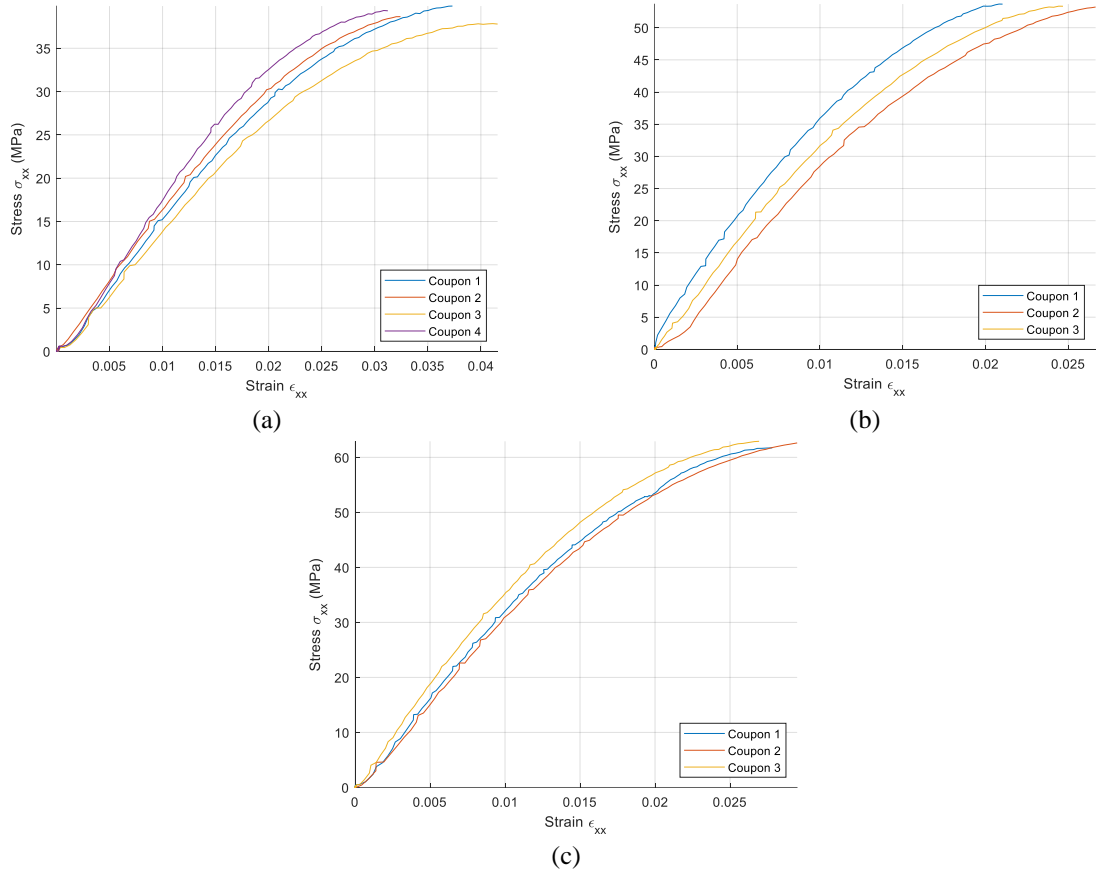

**Figure 6. Full-scale stress-strain curves for the compression tests. (a) Resin. (b) CF-PC. (c) PA12-F30.**

the 3D printed parts. The CF-PC specimens showed higher tensile strength than the 90° tensile strength of T700G/2510.

**B. Mode-II interlaminar fracture properties**

Unfortunately, mode-II interlaminar fracture could not be observed in all the ENF specimens. This was caused by the large ductile behaviors of the specimens. The fracture mode was easily switched from shear to compressive failure due to large bending. Additionally, it was observed that the notch was not perfectly debonded across the midplane. More improvements will be made for the ENF specimens in future work.

## V. Conclusion

This project was focused on characterizing the elastic material properties of specimens manufactured using the latest additive manufacturing techniques. Compared to the 0° properties of the oven-cured T700G/2510 composites,

**Table 4. Test summary and comparison with the 90° properties**

| Test types | | Modulus (GPa) | | | Strength (MPa) | | |
|---|---|---|---|---|---|---|---|
| | | Test | T700G/2510* | % difference | Test | T700G/2510* | % difference |
| Tension tests | Resin | 1.0523 | 8.41 | -87.5 | 30.7818 | 44.3 | -30.5 |
| | CF-PC | 2.9224 | | -65.3 | 49.2500 | | 11.2 |
| | PA12-F30 | 2.9744 | | -64.6 | 39.5960 | | -10.6 |
| Compression tests | Resin | 1.5123 | 8.48 | -82.2 | 39.0269 | 199 | -80.4 |
| | CF-PC | 2.7679 | | -67.4 | 53.4882 | | -73.1 |
| | PA12-F30 | 3.0830 | | -63.6 | 62.4211 | | -68.6 |

*The values were taken from the manufacturer's report [13].

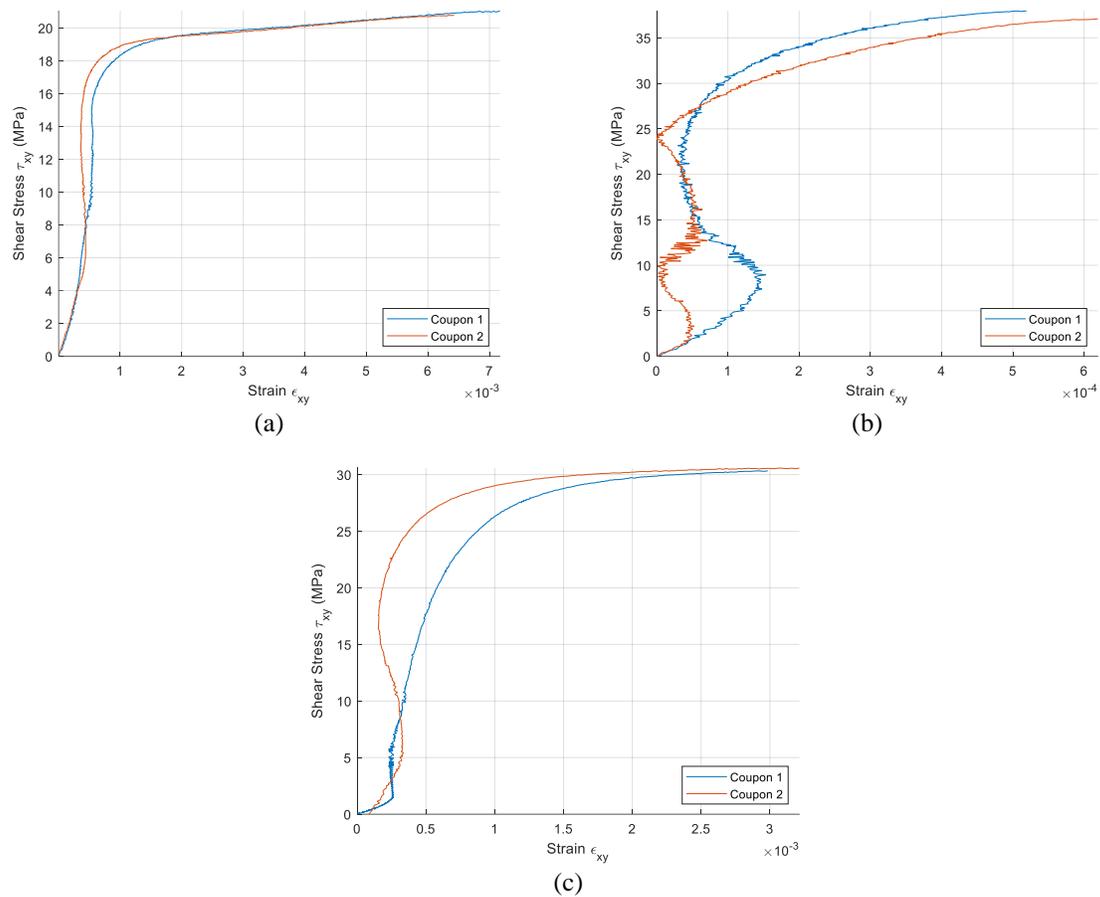

**Figure 7. Full-scale stress-strain curves for the shear tests. (a) Resin. (b) CF-PC. (c) PA12-F30.**

the printed specimens showed significantly smaller modulus and strength values. The specimens showed relatively comparable values compared to the shear and 90° properties of T700G/2510 composites. The CF-PC specimens showed higher tensile strength than the 90° tensile strength of T700G/2510. Future work will be focused on characterizing damage mechanisms in these additive manufacturing materials. This work will contribute to developing additive manufacturing materials and techniques for aerospace and space applications.

## Acknowledgments

The authors greatly thank the support of FAA (Award Number: 12-C-AM-MSU), Jessica Candland in the Department of Aerospace Engineering, Luke Salisbury in the Department of Mechanical Engineering, and Joe Capriotti at the Advanced Composites Institute, Mississippi State University.